\documentclass[12pt]{article}

\begin{document}
\baselineskip=20pt
\date{}
\title{{\bf n+1 dimensional Dirac equation and the Klein paradox}}
\author{\\\\\\\\Antonio S. de Castro\footnote{email:castro@feg.unesp.br}
\\
UNESP/Campus de Guaratinguet\'{a} \\
Caixa Postal 205 \\
12500-000 Guaratinguet\'{a} SP - Brasil \\\\
}
\maketitle

\newpage

In a recent article to this journal, Nitta \textit{et al.} \cite{nitta} have
presented both a derivation of the Dirac equation in 1+1 dimensions and its
solution for the step potential. Furthermore, numerical simulations for the
scattering of a wave packet under the conditions of the Klein paradox are
presented. The purpose of this comment is to clarify two points. First, the
Lorentz structure of the potential and its connection with the Klein
paradox. Second, the connection between the number of space dimensions and
the number of spinor components.

In the Appendix A of Ref. 1 the Dirac equation for a free particle in 1+1
dimensions is derived as it is usually done in the literature for 3+1
dimensions \cite{bd}. The authors found

\begin{equation}
\alpha ^{2}=\beta ^{2}=1,\qquad \alpha \beta +\beta \alpha =0  \label{eq1}
\end{equation}

\noindent concluding that $\alpha $ and $\beta $ are reduced to 2$\times$2
matrices and that any two of the three Pauli matrices can satisfy these
relations. They chose $\alpha =\sigma _{x}$ and $\beta =\sigma _{z}$ and
declared \textit{``In the presence of the scalar potential }$V(x)$,\textit{\
the 1+1 dimensional Dirac equation is extended to the form}

\begin{equation}
\left[ i\hbar \frac{\partial }{\partial t}-V(x)\right] \Psi (x,t)=\left[
c\sigma _{x}\left( -i\hbar \frac{\partial }{\partial x}\right) +\sigma
_{z}m_{0}c^{2}\right] \Psi (x,t)\quad (27)"  \label{eq2}
\end{equation}

\vspace{0.5cm}

\noindent In addition, it is argued that ``\textit{For the case of 2 or 3
dimensions, we have to use the Dirac equation with ordinary \ }4$\times $4 
\textit{Dirac matrices and 4-component spinors because there appears the
spin degree of freedom.'' }In the main body of the paper the authors
presented their calculations for the reflection and transmission amplitudes
(It should be noted in passing that these quantities are indeed amplitudes
but not coefficients as the authors mistakenly stated. Needless to say, the
sum of the coefficients should be equal to one when there exists a
transmitted wave):

\begin{equation}
R=\frac{a-b}{a+b},\qquad T=\frac{2a}{a+b}  \label{eq3}
\end{equation}

\noindent where

\begin{equation}
a=\frac{\sqrt{E^{2}-(m_{0}c^{2})^{2}}}{E+m_{0}c^{2}}\qquad  \label{eq4}
\end{equation}

\begin{equation}
b=\frac{\sqrt{(E-V_{0})^{2}-(m_{0}c^{2})^{2}}}{E-V_{0}+m_{0}c^{2}}
\label{eq5}
\end{equation}

\vspace{0.5cm}

\noindent Thus, they concluded that for $V_{0}>E+m_{0}c^{2}$ there is the
Klein paradox.

The first point to be elucidated is that the potential in the extended form
of the Dirac equation is not a scalar potential as stated by Nitta \textit{%
et al.} Under a Lorentz transformation the potential in Eq. (\ref{eq2})
transforms like the energy,\textit{\ i.e.}, the time component of a Lorentz
vector. On the other hand, a scalar potential should appear in the Dirac
equation multiplied by $\sigma _{z}$ in order to transform itself under a
Lorentz transformation in the same way as the mass of the particle, \textit{%
\ i.e.}, a Lorentz scalar. This would affect Eq. (\ref{eq5}) modifying Eq. (%
\ref{eq4}) by the substitution $m_{0}\rightarrow m_{0}+V_{0}/c^{2}$ instead
of $E\rightarrow E-V_{0}$, leading to no Klein paradox in the presence of a
pure scalar potential. I think it is important to mention that in 1+1
dimensions there are only three linearly independent Lorentz structures for
the potential: scalar, vector and pseudoscalar. This happens because there
are only four linearly independent 2$\times $2 matrices. This \textit{quid
pro quo} between scalar and vector potentials has also appeared recently in
this journal in a paper by Holstein \cite{hol}, where the Klein paradox for
the Klein-Gordon and the Dirac equations was analyzed. In discussing
subbarrier relativistic effects in 3+1 dimensions \cite{anc}, Anchishkin
also unnecessarily regarded the time component of a 4-vector potential as a
scalar potential. It is obvious from the above discussion that erroneous
terms for potentials in relativistic equations may cause confusion to the
unwary.

The second point regards to dimensionality of space. For the generic n+1
dimensions it can be derived that the Hermitian square matrices $\alpha _{i}$
and $\beta $ satisfy the relations $\alpha _{i}^{2}=\beta ^{2}=1$, $\{\alpha
_{i},\beta \}=0$ and $\{\alpha _{i},\alpha _{j}\}=2\delta _{ij}$, where $%
i=1,2,...,$n. It can also be derived that $Tr(\alpha _{i})=Tr(\beta )=0$ and
that their eigenvalues are $\pm 1$, so one can conclude that $\alpha _{i}$
and $\beta $ are even-dimensional matrices. For n$\ =1$ and n $=2$ one can
choose the 2$\times $2 Pauli matrices satisfying the same algebra as $\alpha
_{i}$ and $\beta $, resulting in 2-component spinors in both cases. For n $=3
$ and higher dimensions, though, that is not possible anymore because there
are more matrices required by the algebra than Pauli matrices at one\'{}s
disposal. This is the reason why one has to appeal to 4$\times $4 matrices
and 4-component spinors in 3+1 dimensions. It is true that there is no spin
in the 1+1 dimensional case because there is no angular momentum in one
spatial dimension. Otherwise, in 2+1 dimensions there are only perpendicular
projections of the angular momentum.

\smallskip

\vfill\eject

\end{document}